\documentclass{article}

\usepackage{arxiv}

\usepackage[utf8]{inputenc} 
\usepackage[T1]{fontenc}    
\usepackage{hyperref}       
\usepackage{url}            
\usepackage{booktabs}       
\usepackage{amsfonts}       
\usepackage{nicefrac}       
\usepackage{microtype}      
\usepackage{lipsum}
\usepackage{graphicx}
\usepackage{algorithm} 
\usepackage{algpseudocode}
\usepackage{amsmath}
\graphicspath{ {./images/} }

\title{3D reconstruction of unstained cells from a single defocused hologram}

\author{
 Sunaina Rajora \\
   Department of Physics,\\
 Indian Institute of Technology Delhi,\\
  New Delhi 110016 India. \\
  \texttt{sunainarajora1511@gmail.com} \\
   \And
 Mansi Butola \\
   Department of Physics,\\
 Indian Institute of Technology Delhi,\\
  New Delhi 110016 India. \\
  \texttt{mansibutola83@gmail.com} \\
  \And
 Kedar Khare \\
   Optics and Photonics Center \& \\
   Department of Physics,\\
 Indian Institute of Technology Delhi,\\
  New Delhi 110016 India. \\
  \texttt{kedark@physics.iitd.ac.in} \\
}

\begin{document}
\maketitle
\begin{abstract}
We investigate the problem of 3D complex field reconstruction corresponding to unstained red blood cells (RBCs) with a single defocused off-axis digital hologram. 
We employ recently introduced mean gradient descent (MGD) optimization framework, to solve the 3D recovery problem. While investigating volume recovery problem for a continuous phase object like RBC, we came across an interesting feature of the back-propagated field that it does not show clear focusing effect. Therefore the sparsity enforcement within the iterative optimization framework given the single hologram data cannot effectively restrict the true object volume. For phase objects, it is known that the amplitude contrast of the back-propagated object field at the focus plane is minimum and it increases at the defocus planes. We therefore use this information available in the detector field data to device weights as a function of inverse of amplitude contrast. This weight function is employed in the iterative steps of the optimization algorithm to assist the object volume localization. The experimental illustrations of 3D volume reconstruction of the healthy as well as the malaria infected RBCs are presented. The proposed methodology is simple to implement experimentally and provides an approximate tomographic solution which is axially restricted and is consistent with the object field data.        
\end{abstract}


\section{Introduction}
Holography has historically been associated with 3D imaging capability. In traditional film based holography, the recorded hologram is re-illuminated by conjugate reference beam for replay. The replay field, when viewed by a human observer, provides perception of a 3D image. A critical aspect of this 3D image perception is the ability of human eyes to concentrate on focused objects while ignoring the defocused and blurred background. Film based holography is now largely getting replaced by digital holography where the hologram is recorded on a digital array sensor and the replay process is carried out via numerical computation. The 3D image formation problem in digital holography is quite different in nature. Mimicking the film based holographic replay numerically leads to a reconstructed field in the original object volume, however, now the focusing capability of human eye does not have any role to play in 3D image perception. If the numerically replayed (or back-propagated) object field in the original object volume is carefully examined, one finds that the reconstructed field is quite different from the original object function. In particular, it may be shown that the replay process, at least in the weak scattering approximation, is actually a Hermitian transpose (rather than inverse) operation corresponding to the forward object field formation process at the hologram recording plane \cite{birdi2020}. Since the replay process is linear in nature, there is also a problem associated with degrees of freedom being reconstructed. Formation of a 3D image from a single hologram will mean creation of more information (or voxels) in  a 3D volume out of a smaller number of pixels in the hologram data \cite{birdi2020,javidi2021}. As suggested in the simulation results in \cite{birdi2020}, this degrees of freedom issue may be addressed if the 3D object under consideration is sparse in nature. The 3D image reconstruction problem then may be handled with a sparsity assisted optimization algorithm \cite{birdi2020}. The main aim of the present work is to apply the sparsity based 3D image reconstruction ideas to experimental data consisting of a single defocused hologram corresponding to an unstained red blood cell (RBC) object. Unstained RBCs may be considered as "simple" in their structure, so that, the weak scattering approximation and sparse reconstruction concepts may be applied to this case. In particular it is of interest to know to what extent the 3D object reconstruction by sparse optimization methodology differs from a traditional back-propagation based 3D image recovery.       

Tomographic 3D phase imaging finds applications in number of areas like bio-medical imaging \cite{Tomoreview2021,review2017}, cryo-electron microscopy \cite{cryo2019}, X-ray computed tomography \cite{sidky2008}. The problem of tomographic image reconstruction was first introduced in the early work of Wolf \cite{wolf1969}, where it was showed that the 3D refractive index distribution of an object can be retrieved by solving an inverse scattering problem. This work considered weak scattering approximations to establish a relation between the Fourier transform of the 3D object function and the 2D complex-valued scattered field \cite{wolf1969} as may be recorded using a holographic imaging system. The Fourier relation is known by the name Fourier diffraction theorem \cite{kak2001} and the imaging modality is also sometimes refereed to as optical diffraction tomography (ODT). In a typical ODT setup, multiple holograms of a 3D object are captured at different viewing angles in order to fill the 3D Fourier space of the object as per the Fourier diffraction theorem. 
Advance hardware developments try to record maximum possible number of views either by changing the illumination angle of the incident laser beam on to the sample or by tilting the sample stage itself or by employing both \cite{Tomoreview2021, park2015}. 
In the ODT community more attention has been paid to the problem formulation based on 3D refractive index recovery which inherently arises in the formulation of the Fourier diffraction theorem \cite{kak2001, park2015,Tian2021}. In the present work, we attempt this problem in a slightly different manner where the unknown quantity is the 3D object field function rather than the 3D refractive index distribution, which may make the formulation somewhat simpler. 
Further we use the numerically recovered complex-valued object field in the hologram plane (rather than hologram intensity) as our starting data. The 2D complex object field can be obtained from numerical processing of digital hologram(s) recorded on a digital sensor (CCD or CMOS). The 3D object field to be recovered is considered to be composed of thin slices separated by axial sampling distance $\Delta z$. The phases may then be related to the material present in $\Delta z$ thickness of respective slices. 
In prior literature, different scattering approximations have been suggested like Born or Rytov, based on the phase structure and size of the object \cite{sung2009}. For objects with complex structures where multiple scattering effects can not be neglected, more sophisticated methods like beam propagation method\cite{kamilov2015}, split-step non-paraxial method \cite{Demetri2019} have been suggested. For simplistic objects like human RBCs, first order scattering approximation may be considered sufficient to model the forward problem \cite{park2008,park2013}. In first order Born approximation, each slice of the 3D object volume is propagated independently and the individual propagated fields at the detector plane may be simply added to make the 2D object field. 

Our aim in this work is to investigate the 3D object recovery problem for simple objects from a single defocused hologram recorded on a 2D sensor array. We have studied this problem for simulated discrete objects in \cite{birdi2020}, which we extend here for the real biological samples like RBCs that have continuous structure. Imaging of both healthy and malaria infected RBCs is illustrated. 
The problem of 3D reconstruction with single hologram is studied in literature for objects like particle fields in in-line holographic configuration \cite{Tian2021,tatiana2021}. In \cite{Tian2021}, the beam propagation method is employed to model the forward problem and the 3D refractive index distribution of the particle-field like object function is determined by using a regularized optimization algorithm wherein the measured data is the single in-line hologram intensity. Note that here the role of regularization is to handle both the object twin and defocus components. The work in \cite{tatiana2021volume, tatiana2021} also uses single in-line hologram intensity and iteratively estimates the 3D complex field distribution of discrete particles or text-like objects by application of 3D de-convolution and iterative refinement. We find that for continuous nature of the 3D object under study, a careful examination of the 3D recovery problem is required. For example, if the characteristic of the back-propagated field corresponding to the RBCs is observed, we find that it is quite different from the particle fields. The amplitude of the back-propagated field corresponding to RBCs does not show significant defocusing effect on moving a small distance away from the RBC image plane, as is generally observed for particle fields. With such back-propagated complex fields, the localization of the object volume is thus not a straightforward problem even if the sparsity enforcing priors like total variation are employed in optimization framework. This volume localization problem simplifies for the ODT problem when multiple views of the same sample are captured \cite{krauze2016}. As our present work focuses on the 3D reconstruction with single 2D complex field data at the detector plane, we provide some discussion on the object volume localization in the current problem framework. In particular, we use the inverse amplitude contrast as a weighting factor to improve the confinement of a cell object in the reconstruction volume as we will discuss later in the paper. 
For iterative reconstruction, we employ the optimization methodology of mean gradient descent which was introduced in an earlier work \cite{rajora2019} for single shot interferogram analysis and 2D image de-convolution \cite{rajora2021}. The advantage of this methodology is that it does not require regularization parameter by design and hence the tedious task of empirical tuning is avoided as in standard optimization approaches \cite{goldstein2009,abascal2011,beck2009fast}.

This paper is organized as follows: In Section 2, we describe the problem of 3D volume reconstruction of RBC from a single defocused hologram and further discuss about the peculiar nature of the back-propagated field of the RBC. Section 3 describes the methodology of mean gradient descent optimization and show its application to the problem of 3D reconstruction. In Section 4, we describe the inverse amplitude contrast as a new weighting criteria employed in the iterative optimization algorithm to restrict the object volume. Further we show 3D volume reconstruction results for two examples of a healthy and malaria infected RBCs. Finally in Section 5, we highlight the conclusions and future directions of our work.

\section{Problem overview}
\begin{figure}[tbp] 
\includegraphics[width=\textwidth]{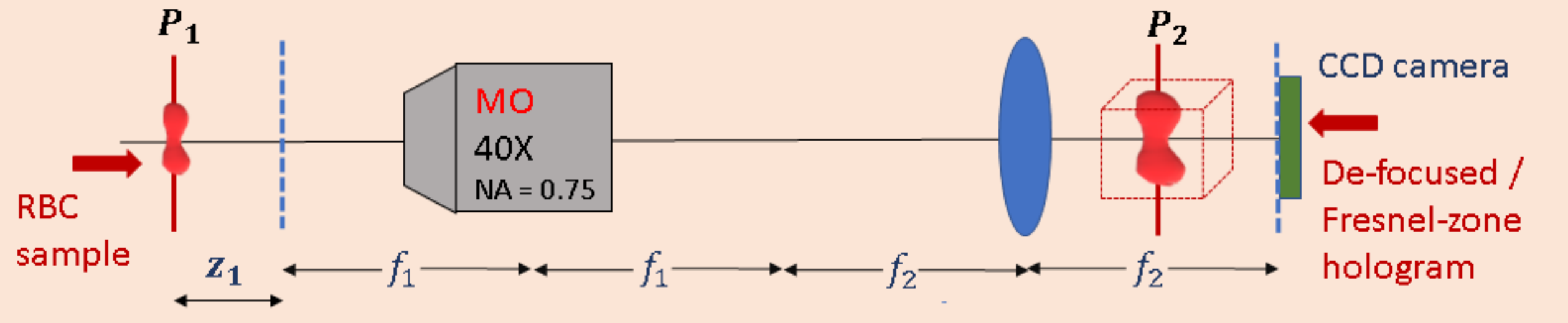}
\caption{System schematic of reconstruction problem. The planes shown with red solid lines correspond to the location of the RBC and its defocused image. Blue dotted lines refer to the perfect focus plane and its conjugate at the detector. Focal lengths $f_1$ and $f_2$ are not shown to scale.} 
\label{fig:setup}
\end{figure}

\begin{figure}
\includegraphics[width=0.75\textwidth]{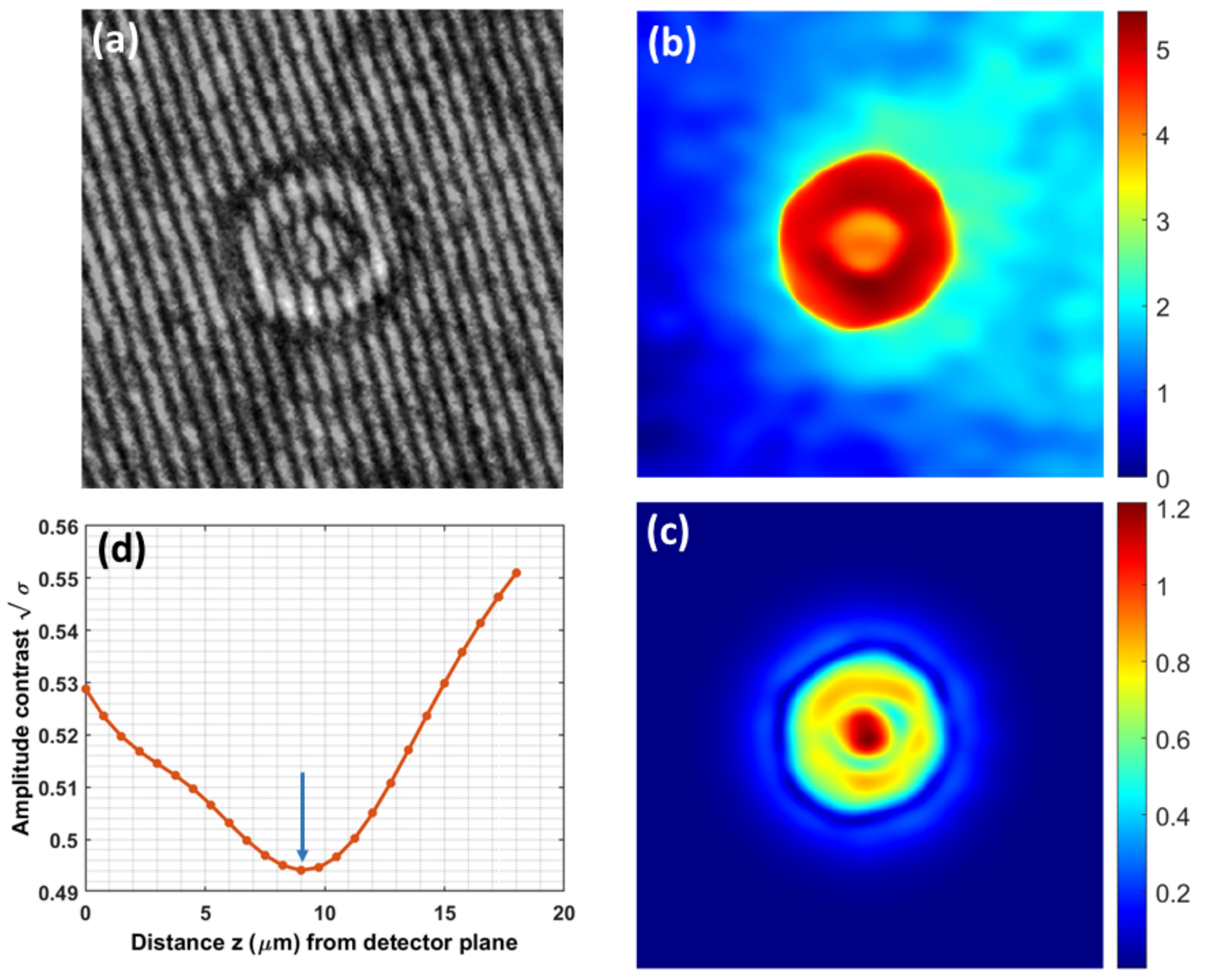}
\caption{(a) ROI of the defocused RBC hologram recorded at 40x magnification in digital holographic microscope. (b)-(c) Amplitude and the unwrapped phase of the 2D complex field $V(x,y)$ reconstructed with FTM. Amplitude is shown after background subtraction and Gaussian windowing. (d) The plot of amplitude contrast metric $\sqrt{\sigma}$ v/s the z-distance from the detector plane. The blue arrow in (d) shows that the amplitude contrast is minimum and hence defocus distance is at 9 $\mu m$.   } \label{fig:sigma}
\end{figure}

The problem we wish to address is illustrated in Fig. \ref{fig:setup}. 
A red blood cell (RBC) object is placed at a defocus plane of an infinity corrected 40x microscopic imaging system which forms one arm of a balanced digital holographic microscope system (Make: Holmarc Product: HO-DHM-UT01-FA). The other arm of the interferometer brings in a tilted plane reference beam. A digital hologram of the defocused cell is recorded on an array sensor (Make: IDS GmBH, uEye 3070, pixel size 3.45 $\mu m$) leading to an off-axis Fresnel zone hologram of the RBC as shown in Fig. \ref{fig:sigma}(a). The DHM system uses a 650 nm diode laser for illumination purposes in the quantitative phase mode. The system is additionally equipped with a switchable white light illumination which makes the system work as an infinity corrected brightfield microscope when required by the users.
To model the problem of 3D reconstruction as in \cite{birdi2020}, the hologram of a single RBC is recorded at a defocus from the perfect focus plane. We identify the focus plane directly in the hologram domain by visually inspecting the fringe pattern. As RBC can be considered essentially as a nearly pure phase object, the image plane hologram predominantly shows phase modulation of fringes (in the form of fringe bending) \cite{Ritika2020}. When the cell is displaced from the focus plane, the phase information is transferred to amplitude and as a result fringes additionally show amplitude modulation in a defocus plane. In the experiment, the defocus is approximately set to $10 \, \mu m$ using a motorized z-stage. The problem of interest now is to use the complex object field recovered in the detector plane and fit it to a 3D object field in a box centered on the location of the image plane of the RBC. 
The 2D complex-valued field at the detector plane is  reconstructed from the off-axis hologram, shown in Fig. \ref{fig:sigma} (a) using Fourier domain filtering of one of the cross-terms in the hologram. Under weak scattering approximation, this complex-valued object field recovered from the recorded hologram can be considered as a sum of the background illumination and the scattered field from the RBC. We remove the effect of background by mean subtraction and further add a Gaussian window over the mean subtracted object field to remove any residual field. The Fourier filtering approach uses the entire hologram, however, we are displaying the resultant phase and amplitude maps of the background subtracted object field in the region of interest (ROI) of $280 \times 280$ pixels (or $24 \mu m$  $\times 24 \mu m$ after accounting for 40x magnification) in Figs. \ref{fig:sigma}(b), (c) respectively. This background subtracted object field is denoted as $V(x,y)$ and is treated as a starting data for our 3D reconstruction problem.
Once we acquired the data $V(x,y)$, now the first step is to locate the defocus distance numerically, which we do by observing the amplitude contrast of the RBC field obtained by back-propagating $V(x,y)$ at different z-distances from the detector plane \cite{autofocusing2008,edge2017}. To find the focus distance, we back-propagate the field $V(x,y)$ via angular spectrum propagation and calculate the standard deviation ($\sigma$) of its amplitude values within a central circular window of radius $65$ pixels. Figure \ref{fig:sigma}(d) shows the plot of amplitude contrast $\sqrt{\sigma}$ against various z-distances from the detector ranging from $0 \mu m$ to $18 \mu m$. We find that the amplitude contrast is minimum at a z-distance of $9 \mu m$, which is indicated by blue arrow in Fig. \ref{fig:sigma}(d). It can be seen that the numerically estimated defocus distance is quite close to the physical displacement ($~10 \mu m$) of the sample stage from the image plane. The defocus estimation is used to define the center of a 3D box within which we wish to reconstruct the object field.
\subsection{Problem model}
We now describe the problem model which will be employed in the later sections to formulate the 3D recovery as a regularized optimization problem. The 3D computational box is as shown in Fig. \ref{fig:setup} extends over $280 \times 280 \times 5$ voxels with physical extent of $24 \times 24 \times 3.75 \mu m^{3}$. As we aim to design the problem in terms of complex fields, our aim is to find the complex fields in each of the 5 slices in the 3D computational box from the knowledge of 2D complex field $V(x,y)$ at the detector plane. The phase distribution in each of the five slices is indicative of the material present in the z-thickness of the respective slices. At this stage we note that the number of unknown voxels is thus five times the number of pixels in the data $V(x,y)$ and such problem can only be addressed by exploiting the object sparsity. 
As RBCs are known to have low structural complexity over the cell volume, as a result, the first order scattering or Born approximation can hold reasonably well \cite{park2013}.
Under this approximation, the background subtracted data $V(x,y)$ may be described in terms of the unknown 3D object field $U(x,y,z)$ which is acted upon by the operator $A$ as follows.
\begin{align}\label{forward}
    V(x,y) &= A U(x',y',z'),   \nonumber  \\
           &= \int_{z_1}^{z_2}dz' \iint dx'dy'\,U(x',y',z') \, h(x-x',y-y', z-z') ,
\end{align}
where the convolution kernel $h$ is defined as the angular spectrum propagator with spatial frequencies $(f_x, f_y)$ defined in accordance with the pixel grid of size $ (3.45/M) \mu m$ with magnification $M = 40$ and band-limited as per the NA (= 0.75) of the microscope objective lens.
\begin{align}
  h(x,y,z) = 
  \mathcal{F}^{-1} \{ \; \exp[i z \sqrt{k^2 - 4\pi^2 (f_x^2 + f_y^2)} \, \cdot \, \textrm{circ}(\frac{\sqrt{f_x^2 + f_y^2}}{(NA/\lambda)}) \; \}.
\end{align}
Note that here we consider the detector placement at the origin (z=0) of the Cartesian coordinate and the 3D computational box is defined at the left half in the z-range $z_1$ to $z_2$ from the detector plane. The forward operation described by $AU$ involves propagating each of the planes in computational volume to the detector plane and adding up the propagated complex-valued fields.   

\subsection{Nature of back-propagated field for a single RBC}
\begin{figure} 
\includegraphics[width=1\textwidth]{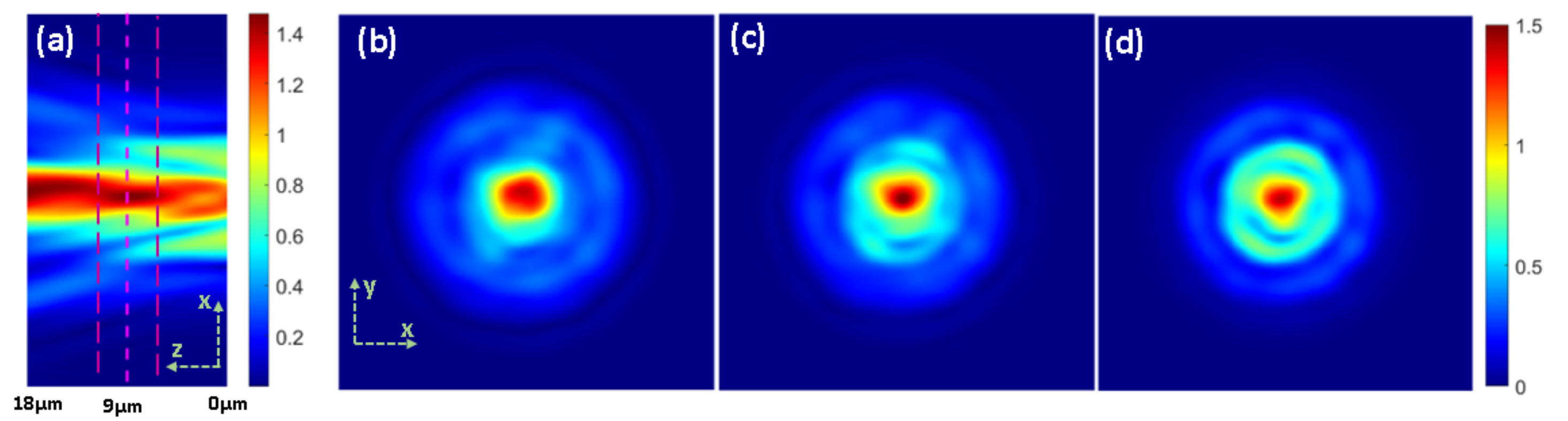}
\caption{(a) The amplitude of back-propagated 2D complex field $V(x,y)$ at various z- distances in the range ( $0-18 \mu m$) shown in the x-z plane, which cuts through the center of y axis. Note that in (a) $0 \mu m$ represents the detector plane. The corresponding amplitude maps in x-y planes located at, (b) 12 $\mu m$, (c) 9 $\mu m$ and, (d) 6 $\mu m$ z-distance from the detector plane as marked by the three dotted magenta lines in (a).  }
\label{fig:nature}
\end{figure}

Before moving on to the next step of solving the 3D recovery problem, it is important to study the nature of back-propagated field corresponding to a single unstained RBC. This observation can provide necessary guidelines for further sections on 3D volume reconstruction. 
Figure \ref{fig:nature}(a) shows the amplitude variation (in x-z plane, which cuts through the center pixel of y axis) of the back-propagated object field in the z-range $0$ to $18 \mu m$ from the detector plane. It can be clearly seen from Fig. \ref{fig:nature}(a) that there is no distinct sign of focusing and defocusing observed in the amplitude of back-propagated field. This is unlike the case of particle field holograms where the back-propagated field amplitude usually shows a focusing behaviour at the location of each of the particles \cite{birdi2020}. The distance between the detector and the central section of RBC volume is already estimated to be $9 \mu m$ in the previous subsection. Therefore, we also observe the amplitude maps (laterally in x-y plane) within $6 \mu m$ z-range around the focus position and find that the amplitude does not get defused as one moves away from the detector plane. The three dotted lines starting from left most in Fig. \ref{fig:nature}(a) are showing the z-positions of the lateral amplitude maps as shown in Figs. \ref{fig:nature}(b), (c) and (d) respectively. 
Owing to this peculiar characteristic of the back-propagated field, the sparsity priors like soft thresholding and total variation (TV) alone may find it difficult to confine the true z-extent of the 3D object within the iterative optimization framework using one-view object field data. Note that in the literature of tomographic reconstruction, a series of holograms are obtained by multiple view sample illumination. Multiple views greatly aid the TV based iterative algorithms to correctly estimate the z-extent of the 3D object \cite{krauze2016}. However, for a single-view hologram based 3D recovery, especially for continuous phase objects, limiting the object extent in this manner does not seem feasible. As we will discuss in Section 4, the amplitude contrast information available in the object field data (see Fig. 2(d)) will prove to be helpful for this purpose.

\section{Methodology}
In this section we present the methodology of the algorithm that we employ to investigate the problem of 3D reconstruction of RBC volume. We start by defining the data inconsistency objective, which quantifies the mismatch between the estimated and known 2D complex field data ($V(x,y)$), and can be mathematically defined as,

\begin{equation}\label{Objective1}
    C_1 = ||V(x,y) - A\,\widetilde{U}(x',y',z')||_{2}^{2},
\end{equation}
 
where the forward operator $A$ is defined in Eq. \eqref{forward} and $\widetilde{U}$ is the guess solution. Here $||.||_{2}$ represents Frobenius norm. It is expected that a solution of optimization problem on forward projection should match the data within the noise limits but this is not a sufficient condition for an incomplete data inverse problem. In that case there can be a large set of solutions which on forward projection satisfy the data constraint very well but do not constitute physically meaningful solutions. The problem can be solved meaningfully only in the case where the object of interest is sparse. Sparsity can be employed in either pixel domain or in some other object domain like in gradient basis. For the present work, we employ total variation (TV) as a sparsity enforcing prior,
\begin{equation}\label{Objective2}
   TV =  C_2 = \sum\sum_{all pixels}\sum \sqrt{|\nabla_{x'}\widetilde{U}|^{2} + |\nabla_{y'}\widetilde{U}|^{2} +|\nabla_{z'}\widetilde{U}|^{2}}.
\end{equation}
Note that the sampling along the lateral dimensions can generally be different from that in the axial dimension which we carefully account for in the definition of numerical calculation of gradients. Other sparsity enforcing penalties like Huber can also be used in this framework. The complex derivatives of the objectives $C_1$ and $C_2$ can be calculated as,
\begin{align}
    \nabla_{\widetilde{U}^{*}} C_1 &= -A^{\dagger} (V(x,y) - A\widetilde{U}(x',y',z')),\\
    \nabla_{\widetilde{U}^{*}} C_2 &= -\nabla . \bigl[ \frac{\nabla \widetilde{U}(x',y',z')}{\sqrt{(|\nabla \widetilde{U}(x',y',z')|^{2}+\epsilon^2)}}\bigr].
\end{align}
Here the operator $A^{\dagger}$ represents error back-propagation \cite{birdi2020} and $\epsilon^2$ can be taken as a small positive number in order to avoid division by zeros. The symbol $\nabla$ represents the standard gradient operator in the 3-dimensional coordinate space $(x',y',z')$. For ease of notation in the further discussion, we define two unit vectors corresponding to the functional gradients $\nabla_{\widetilde{U}^{*}} C_1$ and $\nabla_{\widetilde{U}^{*}} C_2$ as follows:
\begin{align}
    \hat{\bf{d}}_1 &= \nabla_{\widetilde{U}^{*}} C_1 / {||\nabla_{\widetilde{U}^{*}} C_1||_{2}},\\
    \hat{\bf{d}}_2 &= \nabla_{\widetilde{U}^{*}} C_2 / {||\nabla_{\widetilde{U}^{*}} C_2||_{2}},\\
\end{align}
The success of any regularized optimization based object recovery is dependent on the balance between the two objectives $C_{1}$ and $C_{2}$ corresponding to data inconsistency and regularization functionals as defined in Eqs. \eqref{Objective1} and \eqref{Objective2} respectively. The standard approach in such situations is to optimize for a combined cost function of the form $(C_1 + \alpha C_2)$ and obtain a suitable solution by empirically adjusting the regularization parameter $\alpha$. This empirical tuning of $\alpha$ is a tedious task. In an earlier work, we proposed a novel optimization framework that we refer to as mean gradient descent (MGD) which does not require a regularization parameter \cite{rajora2019} by design. The main aim of MGD is to arrive at a solution point where the steepest descent directions $-\hat{\bf{d}}_1$ and $-\hat{\bf{d}}_2$ for the two objectives nearly oppose each other. MGD is inspired by another algorithm ASD-POCS algorithm in the computed-tomography literature \cite{sidky2008} and achieves this equilibrium point by progressing the guess in the mean direction that bisects the unit vectors $\hat{\bf{d}}_1$ and $\hat{\bf{d}}_2$. In particular, the $n^{th}$ iteration of MGD consists of the following procedure:
\begin{equation}\label{MGDiter}
    \widetilde{U}_{n} = \widetilde{U}_{n-1} - t\,(\hat{\bf{d}}_1 + \hat{\bf{d}}_2)_{n-1} /2.
\end{equation}
Here $t$ is the step size for solution update which may be selected using the strategy explained in \cite{rajora2019}. Note that as $\hat{d}_1$ and $\hat{d}_2$ are unit vectors, the solution change due to the two functional gradients is exactly equal in every iteration. An interesting manifestation of balance between the objectives in MGD can be seen in the behaviour of angle $\theta$ between the two steepest descent directions ($-\hat{\bf{d}}_1$ and $-\hat{\bf{d}}_2$),
\begin{equation}\label{Angle}
    \theta = \arccos(\langle{-\hat{\bf{d}}_1,-\hat{\bf{d}}_2}\rangle)
\end{equation}
As the MGD iterations progress, it is observed that the angle $\theta$ starts rising and becomes a large obtuse angle \cite{rajora2019}. At this point the descent directions $-\hat{\bf{d}}_1$ and $-\hat{\bf{d}}_2$ start opposing each other and as they are unit vectors, there is no effective change in the guess solution as per Eq. \eqref{MGDiter} and hence the iterations are stopped. This procedure is able to arrive at a point of equilibrium between the two objectives without the need of any empirically tuned parameter like $\alpha$. 
\section{RBC volume reconstruction with MGD optimization}
\begin{figure}
\caption{Amplitude maps in each of the 5 planes of the reconstructed RBC volume corresponding to the, (a)-(e) back-propagated 3D complex field ($U_B$), (f)-(j) optimization reconstruction $U_{O}$, and (k) -(o) the solution $U_{OW}$ obtained with inverse amplitude contrast ($1/\sqrt{\sigma}$) as a weight metric in MGD iterations. Sideways view in x-z plane of the amplitudes corresponding to the complex fields, (p) $U_B$, (q) $U_O$ and, (r) $U_{OW}$. Note that each row is represented by a common scale bar. } \label{fig:amplitudes}
\includegraphics[width=1\textwidth]{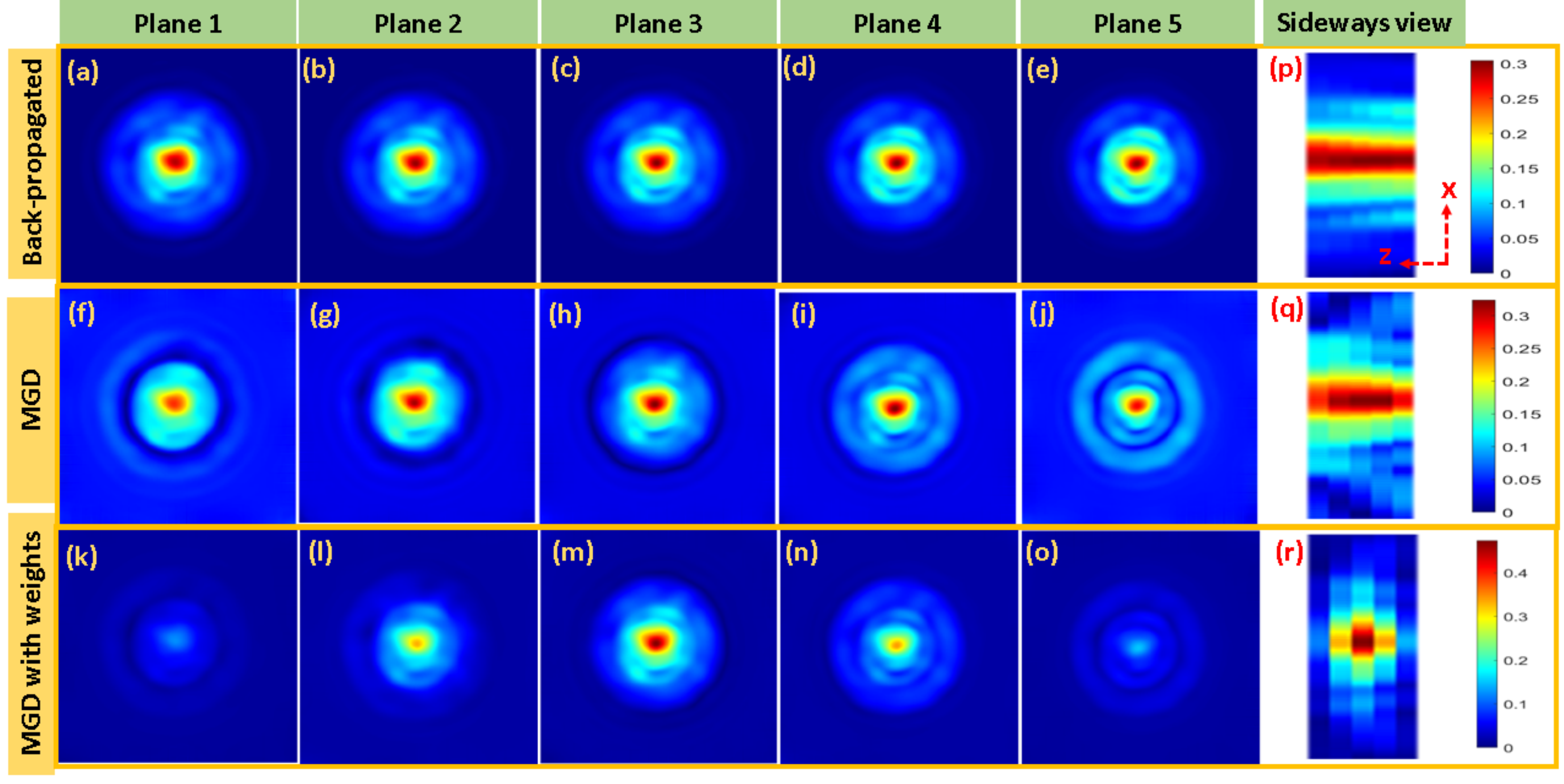}
\end{figure}

\begin{figure}
\caption{Unwrapped phase maps in each of the 5 planes of the reconstructed RBC volume corresponding to, (a)-(e) back-propagated 3D complex field $U_B$, (f)-(j) optimization reconstruction $U_{O}$, and (k) -(o) the solution $U_{OW}$ obtained with inverse amplitude contrast ($1/\sqrt{\sigma}$) as a weight function in MGD iterations. A common scale bar is used to display all the phase maps. } \label{fig:phase}
\includegraphics[width=1\textwidth]{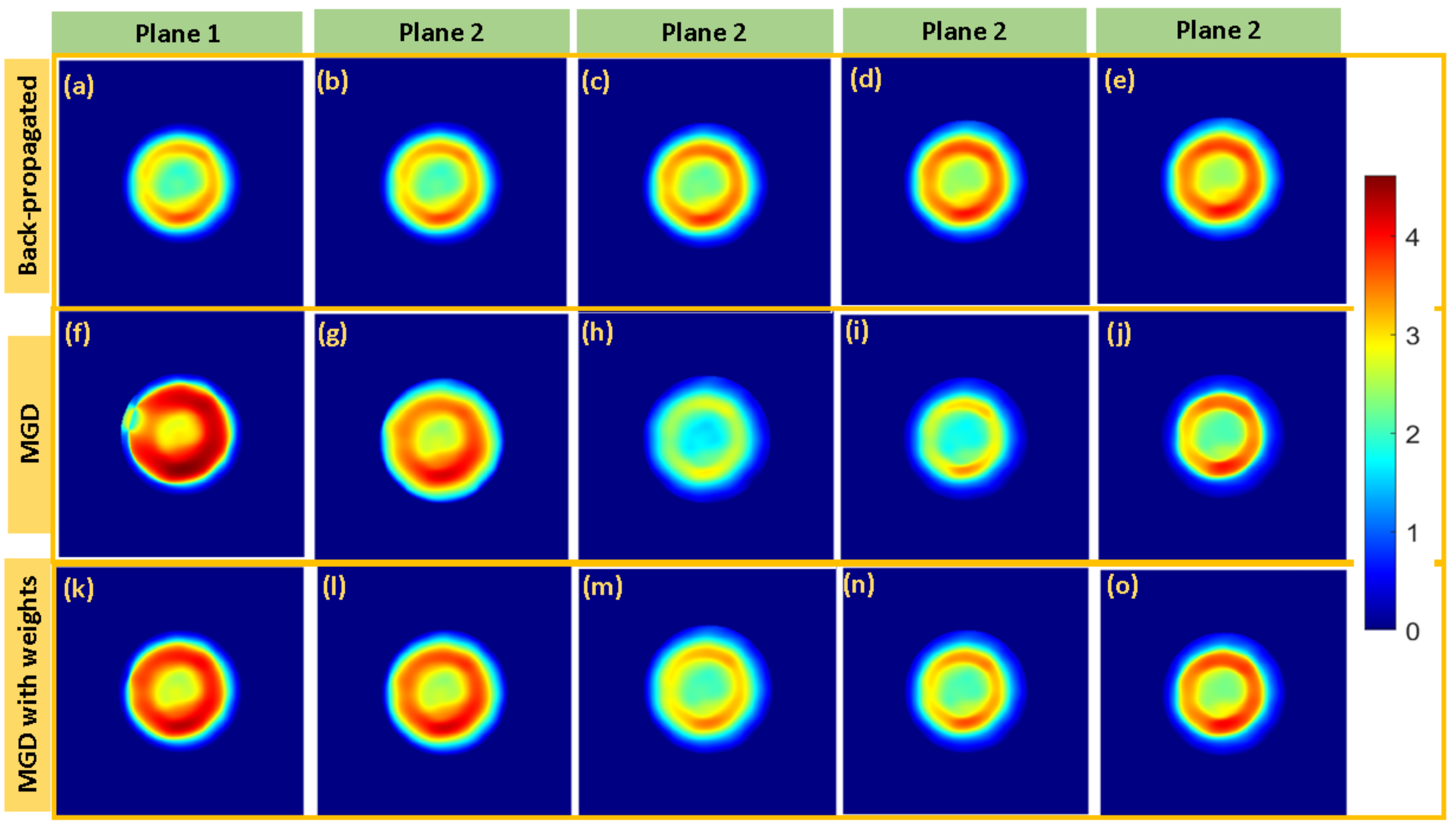}
\end{figure}

Referring to the Sections 2 and 3, we model the problem of 3D complex field reconstruction corresponding to an RBC in the iterative optimization framework of MGD. We denote the back-propagated field over the computational volume of interest generated using the detector field $V(x,y)$ (as shown in Figs. \ref{fig:sigma} (b), (c)) by the symbol $U_B (x,y,z)$. As discussed in \cite{birdi2020}, the back-propagated field can be associated with the operation $\hat{A}^{\dagger}$ as follows:
\begin{equation}
    U_B (x,y,z) = \hat{A}^{\dagger} V(x,y).
\end{equation}
We note that, if $U_B$ is divided by the number of slices in the computational volume, followed by forward propagation (using Eq. \eqref{forward}) of each of the slices to the detector plane, then the resultant integrated object field will exactly match with $V(x,y)$. We therefore use $U_B (x,y,z)/ N_z$ (with $N_z$ equal to number of slices in the volume) with complex-valued noise added to its voxels as the initial guess for the MGD iteration. The first and second rows of Figs. \ref{fig:amplitudes} and \ref{fig:phase} show amplitude and phase maps across the 5 planes of the RBC volume, for the solutions obtained by simple back-propagation ($U_{B}$), and with MGD optimization ($U_{O}$) respectively. It can be clearly seen that for the back-propagation solution $U_B$, there is no appreciable change in the amplitude and phase structure within the reconstruction volume as shown in Figs. \ref{fig:amplitudes} (a)-(e) and \ref{fig:phase} (a)-(e) respectively. Whereas in the optimization solution $U_{O}$, the amplitudes and phases (Figs. \ref{fig:amplitudes} (f)-(j) and \ref{fig:phase} (f)-(j) respectively) do show variations across the reconstructed planes. The relative data domain error $E_d$ for the optimization solution defined as
\begin{equation}\label{Error}
    E_d = \frac{||V(x,y) - A\,\widetilde{U}_O(x',y',z')||_{2}^{2}}{||V(x,y)||_{2}^{2}},
\end{equation}
is observed to be of the order $10^{-8}$ after 500 MGD iterations, suggesting that this solution matches very well with the data. However, we see that, the solution $U_{O}$ still does not exhibit effective axial localization of RBC within the computation volume. We have shown the central x-z slices of the amplitudes $|U_{B}|$ and $|U_{O}|$ in Fig. \ref{fig:amplitudes} (p) and (q) respectively, to highlight this point. 
In this sense the sparsity assisted solution is still not what is desirable.
To address this problem, we now add a weight function depending on the amplitude contrast ($\sqrt{\sigma}$) as in Fig. 2(d) in the iteration process as described next.   

\subsection{Use of amplitude contrast as a weight in iterative reconstruction}
From the previous section, we have observed that the sparsity enforcement alone in the optimization framework, is not sufficient for restricting the axial extent of the RBC. We hereby propose a new weight function that can serve as an additional \textit{a priori} object information and can be used in iterative reconstruction steps to effectively restrict the object volume. From Section 2, we have observed that for phase object like RBC, minimum amplitude contrast ($\sqrt{\sigma}$) serves as an image plane/ focus locating criterion. In addition to detection of defocus distance, the inverse of amplitude contrast can be used as an important information about the true object volume. As illustrated in Fig. \ref{fig:sigma} (d), $\sqrt{\sigma}$ increases with increase in defocus distance, therefore the inverse of $\sqrt{\sigma}$ should nominally act as an appropriate weight function for our purpose. This z-dependent weight function can be used to modify the guess solution within each MGD iteration. In our implementation, we first compute the metric $\sqrt{\sigma}$ for each of the five planes of the back-propagated RBC field ($U_{B}$). An inverse amplitude contrast metric denoted as $w_{\sigma} = 1 / \sqrt{\sigma}$ is computed to be $[0.994, 0.999, 1, 0.999, 0.995 ]$ corresponding to $[1,2,...,5]$ planes respectively. The initial guess for MGD iterations is again taken as the 3D complex field $U_B$ with a small random noise added to it. Figures \ref{fig:amplitudes} and \ref{fig:phase} (k)-(o) show the amplitude and phase maps across the five planes of the reconstructed RBC field, denoted as $U_{OW}$, when the weight function is employed within each of the iterations. Compared to the back-propagated solution shown in Figs. \ref{fig:phase} (a)-(e), the reconstructed phase maps in Figs. \ref{fig:phase} (k)-(o) show appreciable phase variation across each of the planes which is indicative of the different amount of material present in the respective slice thickness.  As can be observed in Figs. \ref{fig:amplitudes} (k) and (o), there is no appreciable amplitude in the peripheral planes of the reconstructed volume. 
Therefore, only three central planes (2,3, and 4) shown in Figs. \ref{fig:amplitudes} and \ref{fig:phase} (l)-(n), when forward projected, effectively contribute to the 2D complex field data $V(x,y)$. This effective axial extent of the RBC volume spans over the physical distance of $~2.25 \mu m$. The axial extent confinement of RBC can also be more clearly seen in the sideways view (x-z plane) of the reconstructed amplitude as shown in Fig. \ref{fig:amplitudes} (r). The relative data domain error $E_d$ (see Eq. \eqref{Error}) between the forward projected 3D reconstruction $U_{OW}$ and the complex field $V(x,y)$ is of the order $10^{-3}$. We want to emphasize that the amplitude contrast information is already available in the data $V(x,y)$ and we have used it explicitly for localization of the cell within the 3D computational volume. As a result the restriction of object field to a limited depth range becomes possible even when a single-view hologram is used as data.
\subsection{Volume reconstruction results for RBC with Malaria parasite}
\begin{figure} 
\caption{(a) Bright-field image of the normal RBC and the RBC infected with malaria parasite as indicated by red arrow, both imaged in the same ROI. (b) The corresponding hologram recorded at 40x magnification on DHM. (c) Amplitude and (d) phase maps of the 2D complex field $V(x,y)$ at the detector plane. } \label{fig:RBC2}
\includegraphics[width=1\textwidth]{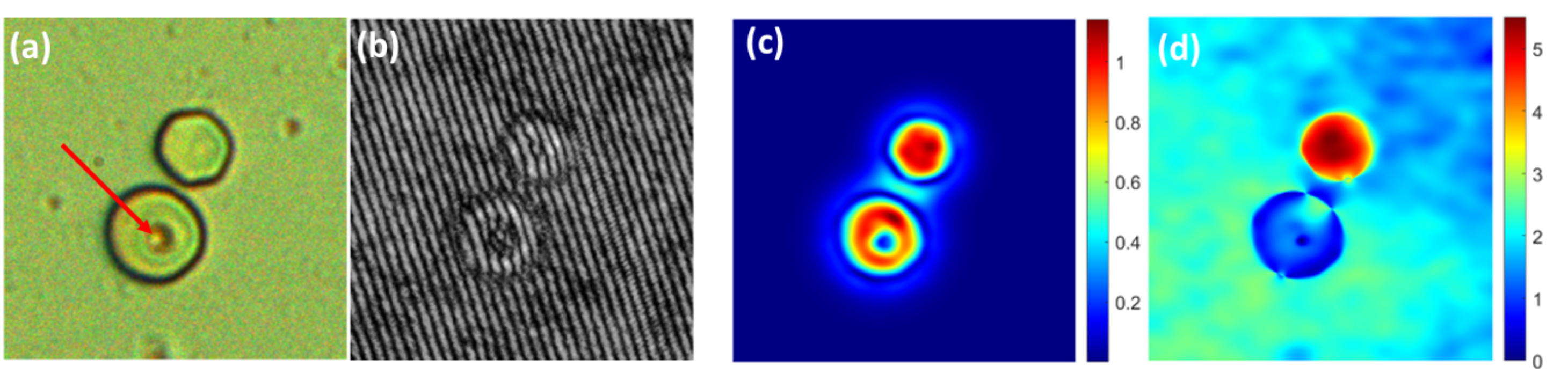}
\end{figure}

\begin{figure} 
\caption{(a)-(e) Amplitude, and (f)-(j) phase maps of the back-propagated field $U_{B}$ across 1-5 planes of RBC volume. (k)-(o) Amplitude, and (p)-(t) phase maps of the complex field $U_{OW}$, obtained by weights assisted optimization, across 1-5 planes of reconstructed RBC volume. The red arrow in (q) shows the phase dip at the location of parasite. } \label{fig:RBC2n}
\includegraphics[width=1\textwidth]{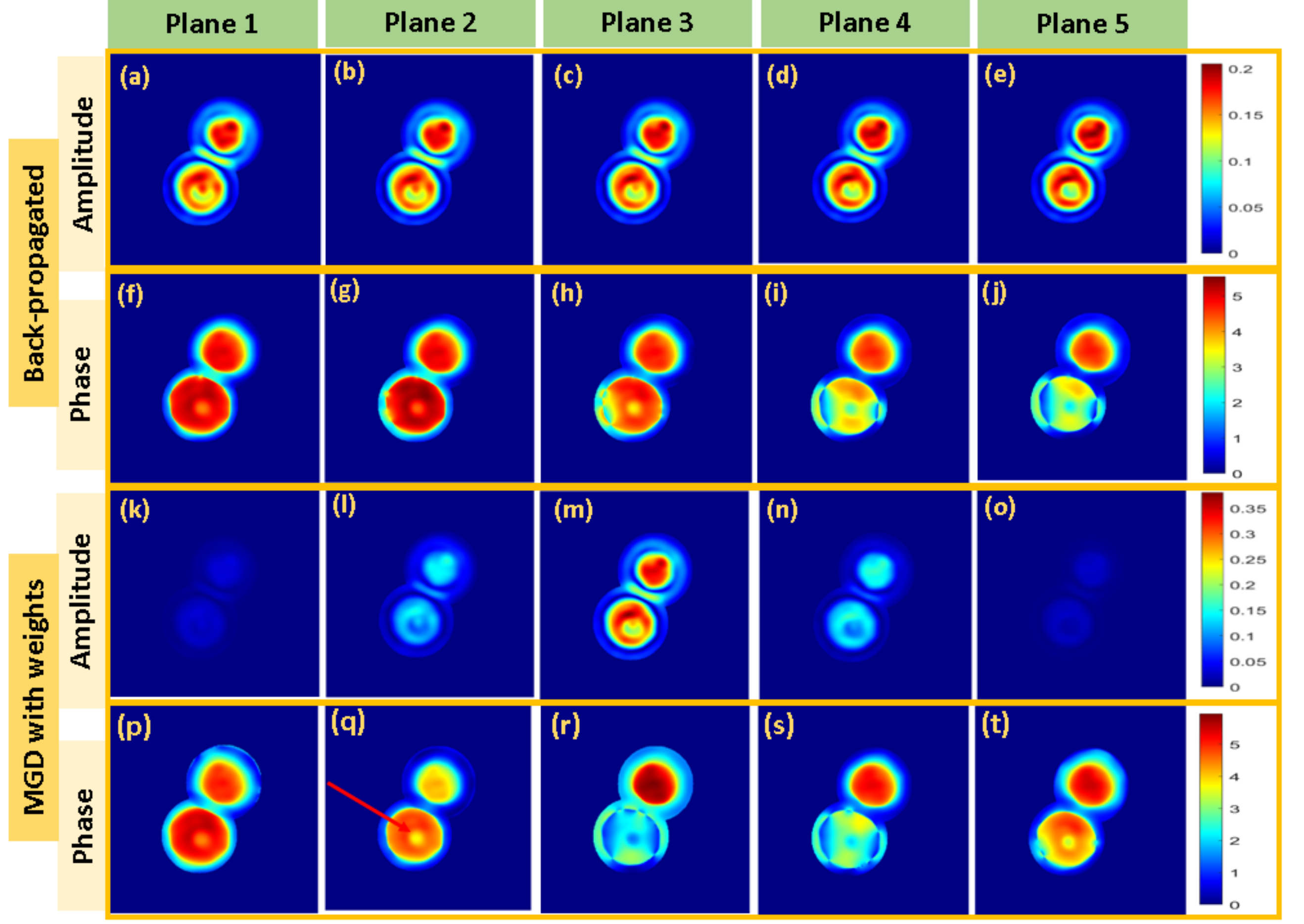}
\end{figure}

The iterative methodology for 3D field reconstruction from a single-view defocused hologram was presented in the context of a single unstained RBC up to this point. In order to validate this methodology, we used the identical algorithm for a malaria infected RBC sample. In particular we selected a ROI of a defocused hologram which contained two adjacently located RBCs, one of which was infected with malaria. To prepare the sample slide of RBC with malaria, the parasites (Plasmodium Falciparum) were cultured in the human RBCs by maintaining appropriate growth conditions for the parasites as described in \cite{trager1976human}. The Plasmodium culture was diluted 1:10 times with 1x PBS and the prepared sample was smeared on a glass slide.
The defocused hologram of the RBC sample is recorded with the same experimental configuration as described in Fig. \ref{fig:setup}. Figures \ref{fig:RBC2} (a) and (b) show the bright-field (BF) image of the cell ROI and the corresponding hologram respectively. The ROI constitutes the computational window of size $360 \times 360$ pixels. As can be clearly seen in the BF image, the selected ROI contains one normal cell which we denote as RBC1 and the other cell, indicated by the red arrow in Fig. \ref{fig:RBC2} (a), which hosts the parasite and is denoted as RBC2. The 2D complex field at detector plane ($V(x,y)$) is computed using Fourier transform method (FTM) and corresponding amplitude and phase maps are shown in Figs. \ref{fig:RBC2} (c), (d) respectively. We calculate the defocus distance by back-propagating the 2D complex field at detector plane due to RBC1, at various z-distances and by using the minimum amplitude contrast criterion as before. The defocus distance from the detector plane is obtained as 4.7 $\mu m$, which is used to define the computational box of size $360 \times 360 \times 5 $ voxels with its center located at 4.7 $\mu m$ away from the detector plane. For this illustration as well, we take random noise added back-propagated field ($U_{B}$) in the computational volume as an initial guess in the MGD iterations. As we have already illustrated in the previous section that in case of single-shot 3D recovery, amplitude contrast based weight function plays an important role in limiting the object extent, therefore we use weights assisted optimization for 3D reconstruction. The computed weights for the present example are $[0.985, 0..996, 1.0, 0.996, 0.982]$. The amplitude and phase maps of the reconstructed RBC field obtained by back-propagation ($U_B$) are shown in Figs. \ref{fig:RBC2n} (a)-(e) and (f)-(j) respectively, and that obtained by employing weights in optimization algorithm, denoted as $U_{OW}$ are shown in Figs. \ref{fig:RBC2n} (k)-(o) and (p)-(t) respectively. 
We observe from Figs. \ref{fig:RBC2n} (a)-(e), that for the field $U_B$, the amplitude is appreciable in all the planes, whereas 
the field $U_{OW}$ (see Figs. \ref{fig:RBC2n} (k)-(o)) shows cell localization along the axial direction. It can also be clearly observed from the phase maps of the reconstructed volume that while the normal RBC (RBC1) has nearly flat phase structure, the host RBC (RBC2) shows a prominent dip in the phase value as shown by red arrow in Fig. \ref{fig:RBC2n}(q), at the location of parasite marked by red arrow in the BF image (see Fig. \ref{fig:RBC2}(a)). We can clearly observe that the phase structure shows variation across the recovered planes shown in Figs. \ref{fig:RBC2n}(p)-(t). 
The relative data domain error ($E_d$) between the forward projected 3D reconstruction $U_{OW}$ and the complex field $V(x,y)$ is of the order of $10^{-2}$ in this case.  

\section{Conclusions}
In summary, we have explored the problem of 3D reconstruction of phase objects from a single-view defocused hologram record. We formulated the 3D imaging problem directly in terms of the complex object field rather than in terms of the 3D refractive index profile. For a continuous phase object (like single RBC), we observed that the back-propagated (or replay) object field to the original 3D volume of interest did not show clear focusing effects. The 3D reconstruction from single-view hologram was then formulated as a sparsity based optimization (with TV penalty). However, it was observed that due to the nature of back-propagated field, the sparsity penalty alone was not successful at restricting the object field axially. It is known that the amplitude contrast of the back-propagated object field for a phase object is minimal in the focus plane. We therefore used a weighting function proportional to the inverse amplitude contrast for different axial slices in the iterative framework. This amplitude contrast information which is contained in the object field data is seen to successfully restrict the 3D object to appropriate axial extent. Further we observe that compared to the simple back-propagation solution, the sparsity based solution with weights shows appreciable change from slice-to-slice in the reconstructed volume. Experimental 3D reconstructions for both normal and malaria infected RBCs are shown in this work for illustrating our proposed approach. We believe that our single-view tomographic reconstruction methodology is simple to implement experimentally and can provide an approximate 3D solution that is axially restricted and is also consistent with the complex-valued object field in the hologram plane. The observations from this study may be useful for multi-view phase tomography systems as well. 

\section*{Acknowledgement}
We would like to acknowledge Dr. Pawan Malhotra and Asif Akhtar, Malaria group, ICGEB, New Delhi for providing samples of malaria infected RBCs.


\begin{thebibliography}{1}
\bibitem{birdi2020}
J. Birdi, S. Rajora, M. Butola, and K. Khare, “True 3d reconstruction in digital holography,” J. Phys. Photonics 2,
044004 (2020).

\bibitem{javidi2021}
B. Javidi, A. Carnicer, A. Anand, G. Barbastathis, W. Chen, P. Ferraro, J. Goodman, R. Horisaki, K. Khare,
M. Kujawinska et al., “Roadmap on digital holography,” Opt. Express 29, 35078–35118 (2021)

\bibitem{Tomoreview2021} 
V. Balasubramani, A. Kuś, H.-Y. Tu, C.-J. Cheng, M. Baczewska, W. Krauze, and M. Kujawińska, “Holographic
tomography: techniques and biomedical applications,” Appl. Opt. 60, B65–B80 (2021).

\bibitem{review2017} 
D. Jin, R. Zhou, Z. Yaqoob, and P. T. So, “Tomographic phase microscopy: principles and applications in bioimaging,” J. Opt. Soc. Am. B 34, B64–B77 (2017).

\bibitem{cryo2019} 
R. Danev, H. Yanagisawa, and M. Kikkawa, “Cryo-electron microscopy methodology: current aspects and future
directions,” Trends Biochem. Sci. 44, 837–848 (2019).

\bibitem{sidky2008} 
E. Y. Sidky and X. Pan, “Image reconstruction in circular cone-beam computed tomography by constrained,
total-variation minimization,” Phys. Medicine \& Biol. 53, 4777 (2008).

\bibitem{wolf1969}  
E. Wolf, “Three-dimensional structure determination of semi-transparent objects from holographic data,” Opt.
Commun. 1, 153–156 (1969).

\bibitem{kak2001} 
A. C. Kak and M. Slaney, Principles of computerized tomographic imaging (SIAM, 2001).

\bibitem{park2015}  
J. Lim, K. Lee, K. H. Jin, S. Shin, S. Lee, Y. Park, and J. C. Ye, “Comparative study of iterative reconstruction algorithms for missing cone problems in optical diffraction tomography,” Opt. Express 23, 16933–16948 (2015).

\bibitem{Tian2021}  
H. Wang, W. Tahir, J. Zhu, and L. Tian, “Large-scale holographic particle 3d imaging with the beam propagation
model,” Opt. Express 29, 17159–17172 (2021).

\bibitem{sung2009}  
Y. Sung, W. Choi, C. Fang-Yen, K. Badizadegan, R. R. Dasari, and M. S. Feld, “Optical diffraction tomography for high resolution live cell imaging,” Opt. Express 17, 266–277 (2009).

\bibitem{kamilov2015}  U. S. Kamilov, I. N. Papadopoulos, M. H. Shoreh, A. Goy, C. Vonesch, M. Unser, and D. Psaltis, “Learning approach to optical tomography,” Optica 2, 517–522 (2015).

\bibitem{Demetri2019} 
J. Lim, A. B. Ayoub, E. E. Antoine, and D. Psaltis, “High-fidelity optical diffraction tomography of multiple scattering,” Light. Sci. \& Appl. 8, 1–12 (2019).

\bibitem{park2008}  Y. Park, M. Diez-Silva, G. Popescu, G. Lykotrafitis, W. Choi, M. S. Feld, and S. Suresh, “Refractive index maps and membrane dynamics of human red blood cells parasitized by plasmodium falciparum,” Proc. Natl. Acad. Sci. 105, 13730–13735 (2008).

\bibitem{park2013} 
K. Kim, H. Yoon, M. Diez-Silva, M. Dao, R. R. Dasari, and Y. Park, “High-resolution three-dimensional imaging of red blood cells parasitized by plasmodium falciparum and in situ hemozoin crystals using optical diffraction tomography,” J. Biomed. Opt. 19, 011005 (2013).

\bibitem{tatiana2021}  
T. Latychevskaia, “Three-dimensional structure from single two-dimensional diffraction intensity measurement,” Phys. Rev. Lett. 127, 063601 (2021).

\bibitem{tatiana2021volume} 
T. Latychevskaia, “Three-dimensional volumetric deconvolution in coherent optics and holography,” Appl. Opt. 60, 1304–1314 (2021).

\bibitem{krauze2016} 
W. Krauze, P. Makowski, M. Kujawińska, and A. Kuś, “Generalized total variation iterative constraint strategy in limited angle optical diffraction tomography,” Opt. Express 24, 4924–4936 (2016).

\bibitem{rajora2019} 
S. Rajora, M. Butola, and K. Khare, “Mean gradient descent: an optimization approach for single-shot interferogram analysis,” J. Opt. Soc. Am. A 36, D7–D13 (2019).

\bibitem{rajora2021} 
S. Rajora, M. Butola, and K. Khare, “Regularization-parameter-free optimization approach for image deconvolution,” Appl. Opt. 60, 5669–5677 (2021).

\bibitem{goldstein2009} 
T. Goldstein and S. Osher, “The split bregman method for l1-regularized problems,” SIAM J. on Imaging Sci. 2,
323–343 (2009).

\bibitem{abascal2011}
J.-J. Abascal, J. Chamorro-Servent, J. Aguirre, S. Arridge, T. Correia, J. Ripoll, J. J. Vaquero, and M. Desco, “Fluorescence diffuse optical tomography using the split bregman method,” Med. Physics 38, 6275–6284 (2011).

\bibitem{beck2009fast} 
A. Beck and M. Teboulle, “A fast iterative shrinkage-thresholding algorithm for linear inverse problems,” SIAM J. Imaging Sci. 2, 183–202 (2009).

\bibitem{Ritika2020} 
R. Malik, P. Sharma, S. Poulose, S. Ahlawat, and K. Khare, “A practical criterion for focusing of unstained cell samples using a digital holographic microscope,” J. Microsc. 279, 114–122 (2020).

\bibitem{autofocusing2008}
P. Langehanenberg, B. Kemper, D. Dirksen, and G. Von Bally, “Autofocusing in digital holographic phase contrast
microscopy on pure phase objects for live cell imaging,” Appl. Opt. 47, D176–D182 (2008).

\bibitem{edge2017}
 Y. Zhang, H. Wang, Y. Wu, M. Tamamitsu, and A. Ozcan, “Edge sparsity criterion for robust holographic autofocusing,”
Opt. Lett. 42, 3824–3827 (2017).

\bibitem{trager1976human} 
W. Trager and J. B. Jensen, “Human malaria parasites in continuous culture,” Science 193, 673–675 (1976).
\end{thebibliography}
\end{document}